\documentstyle[12pt]{article}
\begin{document}
%\setlength{\textheight}{20 cm}
%\begin{document}
\begin{titlepage}
\begin{flushright}
TP-USl/95/04
\vspace{1 cm}
\end{flushright}
\begin{center}
{\LARGE\bf On possibility of detecting the $e^-e^- \rightarrow W^-W^-$
process in the Standard Model} \\
\vspace{1.5 cm}
{\large\bf J.Gluza}\footnote{e-mail address: gluza@us.edu.pl}
and {\large\bf M.Zra{\l}ek}\footnote{e-mail address:
zralek@us.edu.pl} \\
\vspace{ 0.5 cm}
Department of Field Theory and Particle Physics \\
Institute of Physics, University of Silesia \\
Uniwersytecka 4, PL-40-007 Katowice, Poland \\
\vspace{1cm}
%\maketitle
\baselineskip 1 mm
{\large\bf Abstract} \\
\end{center}
We examine the $e^-e^- \rightarrow W^-W^-$ lepton number violating process
in the frame of the Standard Model with additional right-handed neutrino
singlets.
We give results in the framework of the `see-saw' as well as other models,
where
there is no relation between neutrino masses and the mixing matrix elements.
The cross section for the `see-saw' models is negligible because they predict
in a natural way very small electron-heavy
neutrino mixing angles. However, there exist other models
in which the electron-heavy mixing angles are free parameters and
can be large. Taking into account the present experimental bounds on mixing
angles the large cross section for the $e^-e^- \rightarrow W^-W^-$
is still acceptable.
\end{titlepage}
\vspace{0.5 cm}
\baselineskip 6 mm
\section{Introduction}

   Construction of high energy electron-electron accelarator is technically
viable [1]. The $e^-e^-$ process is very interesting because it is particularly
suitable for testing a possible lepton violation mechanism. Observation of the
processes such as $e^-e^- \rightarrow \mu^-\mu^-,\;\tau^-\tau^-\;{\rm
or\;}W^-W^-$
will indicate the family lepton number or the total lepton number violation.
Especially interesting is the last process, $e^-e^- \rightarrow W^-W^-$, where
the
total lepton number is violated by two units, $\Delta L=2$. Firstly, it
has been shown
that the standard model's background for such process  can be substantially
reduced
by appropriate kinematical
cuts (e.g. below 0.1 fb for a 1 TeV $e^-e^-$
collider [2,3]). Secondly, the occurence of this process will indicate that
there
exist
massive neutrinos (with masses $M_N>M_Z$) of Majorana type. Several papers
have been devoted to such breaking process with $\Delta$L=2 in the last few
years but
with different conclusions. Some are very pesimistic and
indicate that the total cross section for the $e^-e^- \rightarrow W^-W^-$
process
is much below the SM background [4], others [5], are very optimistic and
predict
that $\sigma_{tot}(e^-e^- \rightarrow W^-W^-)$ can be as large as 4 fb
(for $\sqrt{s}=0.5$ TeV) or 64 fb (for $\sqrt{s}=1$ TeV).

We would like to elucidate this point. Is there realy a chance to observe
such process in the future NLC ($\sqrt{s}=0.5$ TeV) or TLC ($\sqrt{s}=1$ TeV)
colliders? The answer depends on the model in which we calculate
the cross section. We take the simplest one -
the Standard Model (SM) with additional heavy right-handed neutrino singlets
(RHS). In that case
the
process takes place by exchange of neutrinos in the t-and u-channels. But even
then the size of the total cross section depends on the way in which light
and heavy neutrino masses are generated. The number of heavy neutrinos,
the magnitudes of their masses and the high energy behaviour of the total
cross section (unitarity) also have strong consequences.

In Section 2 we give the necessary information about the SM with right-handed
neutrinos, the helicity amplitudes for the $e^-e^- \rightarrow W^-W^-$ process
and various limits of the cross section. In the main section (Section 3)
we discuss the numerical results of our calculations and finally we summarize
and conclude  in Section 4.

\section{The RHS model and the cross section}
%\subsection{Cross section in the frame of the RHS model.}

In the RHS model which we consider there are $n_L$ (=3) left-handed and $n_R$
(=1,2,...) right-handed weak
neutrino states transforming under $SU_L(2)$ gauge group
as doublets and singlets,
respectively. The neutrino mass matrix has $n_L+n_R$
dimensions
\begin{equation}
M_{\nu}= { \overbrace{0}^{n_L} \ \overbrace{M_D}^{n_R} \choose M_D^T \ M_R }
{\begin{array}{c}
\} n_L \\ \} n_R.
\end{array}}
\end{equation}
Without Higgs triplet fields the $n_L \times n_L$ dimension part $M_L$ of
$M_{\nu}$ equals zero
\begin{equation}
M_L=0.
\end{equation}
Using ($n_L+n_R$) dimensional unitary matrix $U=\left( \matrix{ K^T \cr
                                                      U_R } \right) $
acting on the weak neutrino states we can diagonalize matrix $M_{\nu}$
($U^TM_{\nu}U=M_{diag}$) and get the
physical states. We know from experiments that three of
them are very light ($m_{\nu_e}<5.1\;{\rm eV},\;m_{\nu_{\mu}}<270\;{\rm keV},
\;m_{\nu_{\tau}}<24\;{\rm MeV}$)
and others, if exist, have masses above $M_Z/2$ [6] or even $M_Z$ with
appropriate assumptions about their couplings [7].

Without loosing the
generality we can assume that the charged lepton mass matrix is diagonal, so
then the physical neutrino $N= \left( N_1,...,N_{L+R} \right)^T$
couplings to gauge bosons are defined by
($\hat{l}=(e,\mu,\;\tau)^T,\;P_L=\frac{1}{2}(1-\gamma_5)$)
\begin{eqnarray}
L_{CC}&=&\frac{g}{\sqrt{2}}\bar{N}\gamma^{\mu}KP_L\hat{l}W_{\mu}^+ + h.c., \\
L_{NC}&=&\frac{g}{2\cos{\theta_W}} \left[
\bar{N}\gamma^{\mu}P_L(KK^{\dagger})N \right].
\end{eqnarray}
For $n_R=3$ the matrix K
has the following form
$$
K=
\begin{array}{c}
e\;\; \mu \;\; \tau \;\;\;\;\;\;\;\;\;\;\;\;\;\;\;\;\;\;\;\;\;\;\;\;\;\; \\
  \left. { \begin{array}{ccc}
            \cdot & \cdot & \cdot \\
            \cdot & \cdot & \cdot \\
            \cdot & \cdot & \cdot
            \end{array} }
  \right\} {\rm light\;\; neutrinos} \\
   \left. { \begin{array}{ccc}
            \Box & \cdot & \cdot \\
            \Box & \cdot & \cdot \\
            \Box & \cdot & \cdot
            \end{array} }
   \right\} {\rm heavy\;\; neutrinos}
\end{array}.
$$

We are specially interested in the relevant for our process `box' couplings
of electrons with heavy neutrinos.
{}From various experimental data we can find the bounds on the mixing matrix
elements $K_{Nl}$ and $(KK^{\dagger})_{NN'}$ [8,9].

Production of two gauge bosons
in two-charged-electrons scattering process in the SM with only
the RHS neutrinos is described by the helicity amplitudes with the same
negative
polarizations of the incoming electrons
$\sigma_1=\sigma_2=-1/2$. The other helicity
polarizations of electrons are connected with right-handed currents which
are absent - Eqs.(3,4)
(the full helicity amplitudes suitable for the L-R models with additional
s-channels and right handed currents
are given e.g. in [4]).
The RHS model's differential cross section is given by
\begin{equation}
\frac{d\sigma(\lambda_1,\lambda_2)}{d\cos{\Theta}}=
\frac{G_F^2\sqrt{1-\gamma^2} }{ 16 \pi }{\mid M(\lambda_1,\lambda_2) \mid }^2
\end{equation}
where  $\lambda_1$ and $\lambda_2$ are helicities of
the produced gauge bosons and
$\gamma=\frac{2M_W}{\sqrt{s}}$.
The helicity amplitudes can be written in the form ($\Theta,\;\phi$ are polar
angles of one of the gauge bosons in the CM frame)

\begin{equation}
M( {\lambda}_1,{\lambda}_2 ) =
\left\{ M_t(\lambda_1,\lambda_2)R_t
+M_u(\lambda_1,\lambda_2)R_u  \right\}
D^{\mid \lambda_1-\lambda_2 \mid }_{0, \; \lambda_1-\lambda_2} \left(0,
\Theta,\phi \right)
\end{equation}
where $R_t$ and $R_u$ are as follows ($m_a$ - masses of
neutrinos,$\beta=\sqrt{1-
\gamma^2}$)
\begin{eqnarray}
R_{t(u)}  &=& - \sum_{a}
K_{ae}^2
\frac{m_a}
{\frac{1+\beta^2}{2} \mp \beta\cos{\Theta}+\frac{m_a^2}{s}}
\end{eqnarray}
and the
reduced helicity amplitudes $M_{t(u)}\left( \lambda_1,
\lambda_2\right) $ for the t- and u-channels are gathered in Table 1.
The sum $\sum_{a}$ is over all light and heavy neutrinos.
Now we can easily find the approximated cross section formulae in some limited
cases.

(i) For very high energy, if ($\sqrt{s} >> M_W,m_a$) only one helicity
amplitude M(0,0) gives non-vanishing contribution and
\begin{equation}
\sigma(s \rightarrow \infty ) = \frac{G_F^2}{4\pi} \mid
\sum_{a} K^2_{ae} m_a \mid ^2,
\end{equation}
In the RHS model, however,
\begin{equation}
\sum_{a} K^2_{ae}m_a=(M_L^{\ast})_{\nu_e \nu_e}=0,
\end{equation}
so unitarity is restored.

(ii) If additional right-handed neutrinos are very heavy and \newline
$m_{heavy(a)} >> \sqrt{s} >> M_W$ we get
\begin{eqnarray*}
\sigma(m_{heavy(a)}>>\sqrt{s}>>M_W)&=&\frac{G_F^2s^2}{4\pi} \mid
\sum_{light(a)}(K_{ae})^2\frac{m_a}{s}+ \sum_{heavy(a)}
(K_{ae})^2\frac{1}{m_a} \mid ^2 \nonumber \\
&=&\frac{G_F^2}{4\pi} \mid
\sum_{heavy(a)} (K_{ae})^2m_a\left[ 1-\frac{s}{m_a^2} \right] \mid ^2,
\end{eqnarray*}
\begin{equation}
\end{equation}
where we used Eq.(9).
Let us note that  $s/m_a^2 <<1$ for $m_{\rm heavy(a)}>>\sqrt{s}$ and
the contribution from the light neutrinos dominates.

(iii) If masses of the heavy neutrinos $m_{heavy} \rightarrow \infty,$
then
\begin{eqnarray}
\sigma(m_{heavy} \rightarrow \infty )
&=&\frac{G_F^2}{4 \pi}
\mid \sum_{light(a)} (K_{ae})^2m_a \mid ^2 <
\frac{G_F^2}{4 \pi}
(m_{\nu_e}+m_{\nu_{\mu}}+m_{\nu_{\tau}})^2 \nonumber \\
& \sim & \frac{G_F^2}{4 \pi} (25 MeV)^2
\; or \; \frac{G_F^2}{4 \pi} (30\;eV)^2 \nonumber \\
& \sim & 10^{-2} fb \; or \; 10^{-13} fb.
\end{eqnarray}
The first (second) estimation is connected with limits on the light neutrino
masses
coming from terrestrial experiments (astrophysical and cosmological
observations respectively).

We can see from Eq.(11) that the contribution to $\sigma (e^-e^- \rightarrow
W^-W^-)$
from light neutrino exchange is very small ($<10^{-2}$ fb). It follows from
Eq.(10) that the light neutrino dominates if $m_{\rm heavy(a)} >> \sqrt{s}$,
and from the unitarity $\sigma \rightarrow 0$ if $\sqrt{s} >> m_{\rm heavy(a)}$
(Eq.(9)). So the only region where $\sigma (e^-e^- \rightarrow W^-W^-)$
could be large is for $\sqrt{s} \sim m_{\rm heavy(a)}$. We will see in the
next section that it realy takes place.
%%%%%%%%%%%%%%%%%%%%%%%%%%%%%%%%%%%%%%%%%%%%%%%%%%%%%%%%%%%%%%%%%%%%%%%%%%%%%%
\section{The $e^-e^- \rightarrow W^-W^-$ process: numerical results}
%%%%%%%%%%%%%%%%%%%%%%%%%%%%%%%%%%%%%%%%%%%%%%%%%%%%%%%%%%%%%%%%%%%%%%%%%%%%%%

All specific features of our lepton-violating process are included in
quantities $R_t$ and $R_u$ given by Eq.(7). Each of them is expressed by a
sum  over physical neutrinos with masses $m_a$, where the three are light ones
and the others
are heavy. The magnitudes of the $R_t$ and $R_u$ decide about the size of
the total cross section. First of all, if all $m_a \rightarrow 0,\;R_{t(u)}
\rightarrow 0$ as it should be, because for the Weyl neutrinos there
is no lepton symmetry breaking.
Two factors influence the magnitude
of the $R_{t(u)}$ - the square of the mixing matrix elements $K_{ae}^2$ and
neutrino
masses $m_a$ which are restricted by two constraints:
\begin{itemize}
\item unitarity of the K matrix
\begin{equation}
\sum_{i=light} \mid K_{ie} \mid^2=1-\sum_{j=heavy} \mid K_{je} \mid^2
\end{equation}
and
\item the lack of Higgs triplets from which $M_L=0$ and Eq.(9) follows
\begin{equation}
\Delta_{light} \equiv \sum_{i=light}(K_{ie})^2m_i=-\sum_{j=heavy}(K_{je})^2m_j
\equiv -\Delta_{heavy}.
\end{equation}
\end{itemize}

Just above the threshold for the $W^-W^-$ production $\sqrt{s} \sim 2M_W$
and the cross section is small ($\gamma \rightarrow 1$). As we try to find the
region where the total cross section is largest we take $\sqrt{s} >>M_W$,
so $\gamma \rightarrow 0$ and only one helicity amplitude M(0,0) is important,
the other ones tend to zero like $\gamma^2$ or $\gamma$ (see Table 1). We
consider this limit in our
discussion but numerical results are given without any
approximation.

We can ask now what the contribution to $\sigma$ from light neutrinos is.
This contribution is given by Eq.(11), for $m_{heavy} \rightarrow \infty .$
As the masses are small and $\mid K_{ae} \mid \leq 1,$ $\Delta_{light}$ in
Eq.(13) is very
small too and the cross section is $<10^{-2}$ fb or $10^{-13}$ fb for
laboratory
experiments or
astrophysical observations, respectively. The only possibility to obtain larger
cross section is through heavy neutrino(s) exchange in the t- and u-channels.
But even
if the masses of heavy neutrinos are very large the combination
$\Delta_{heavy}=
-\sum\limits_{j=heavy}K_{je}^2m_j$ must be still very small as $\mid
\Delta_{heavy} \mid=
\mid \Delta_{light} \mid .$

The combination $\Delta_{heavy}$ given by the heavy neutrino
exchange (Eq.(13)) can be small because of two reasons. Firstly, the mixing
matrix element $\mid K_{ie} \mid $ can be small so even for large $m_i$
the combination $\Delta_{\rm heavy}$ is small. As we will see this happens
in the case of the `see-saw' type [10] of the neutrino mass matrix $M_{\nu}$.
There is no chance then to get a reasonably large cross section.
Secondly, $\mid K_{ie} \mid $ are not small but there is destructive
interference
between large
contributions from the different heavy neutrinos. It can happen if the
neutrinos
have opposite CP parities.
The models where this scenario is realized are also considered [11].
There is a possibility to get
`experimentally interesting' value of the cross section in the frame of these
models.
The neutrino propagator
gives the factors $+\frac{m_a^2}{s}$ in the denominators of $R_{t(u)}$
(Eq.(7)).
These factors can disturb the destructive interference in $\Delta_{heavy}$
giving larger
values for $R_{t(u)}$.

If the masses of heavy neutrinos are equal then
we can extract the denominator in $R_{t(u)}$ and the cross section is still
proportional to $\mid \Delta_{heavy} \mid^2$ so it is small.
The same is true if there is only one
heavy neutrino. Then the destructive interference in $\Delta_{\rm heavy}$
is impossible and the cross
section $\sigma(e^-e^- \rightarrow W^-W^-)$ is of the same order of magnitude
as for the light neutrinos. The last observation agrees with the fact that in
our model with one heavy neutrino only the `see-saw' scenario is applicable
[12].
{}From the present limits on the masses of
light neutrinos we can see that at least two conditions must be satisfied to
get experimentally
interesting value of the cross section; (i) there must be two or more heavy
neutrinos and (ii) their masses must be different.
We see also that some heavy neutrino mixing angles $K_{ie}$ must be complex,
which can happen if CP is violated or if CP is conserved and the CP parities
of some heavy neutrinos are opposite.

 The spectrum of the neutrino masses and elements of the mixing matrix K are
the
result of the diagonalization of neutrino mass matrix $M_{\nu}$. The elements
of the $M_{\nu}$ are not known and usually some models which guarantee
a reasonable spectrum of neutrino masses are assumed. The popular model
to obtain the light ($\sim$eV) - heavy ($\sim$TeV) spectrum of neutrino masses
is the
`see-saw' model [10]. This means that the $M_R$ and $M_D$ matrices in Eq.(1)
are proportional to different scales of symmetry breaking and
$\mid (M_R)_{ii}\mid >> \mid (M_D)_{lk} \mid $. Then, without any additional
symmetry, the important K matrix elements $K_{ae}$ are proportional to
$<M_D>/m_a$ and are very small for large $m_a$
\begin{equation}
K_{ae} \sim \frac{<M_D>}{m_a}.
\end{equation}
In this case not only $\Delta_{\rm heavy}=-\sum\limits_{a=1}^{n_R}(K_{ae})^2
m_a \simeq \sum\limits_{a=1}^{n_R} \frac{<M_D>}{m_a} $ but also the
quantities $R_{t(u)}$
\begin{equation}
R_{t(u)} \simeq \sum \frac{<M_D>}{m_a \left( \frac{1+\beta^2}{2}\mp
\beta \cos{\Theta}+\frac{m_a^2}{s} \right) }
\end{equation}
are small. The same phenomena of decoupling of the heavy neutrinos in the
`see-saw' type of models at the one-loop level have been also observed (see
e.g. [13]). To find what the size of total cross section is let us take
the neutrino mass matrix in the following `see-saw' forms
\begin{equation}
M_D= \left( \matrix{ 1.0 & 1.0 & 0.9 \cr
                   1.0 & 1.0 & 0.9 \cr
                   0.9 & 0.9 & 0.95 } \right)
\;\;{\rm and}\;\;M_R= \left( \matrix{ M & 0.0 & 0.0 \cr
                      0.0 & AM   & 0.0 \cr
                      0.0 & 0.0 & BM } \right)
\end{equation}
which give a reasonable spectrum of the neutrino masses for M$>$100 GeV,
A,B$>$10
($m_{\rm light} =0 {\rm \;eV},\sim {\rm keV}, \sim {\rm MeV},m_{\rm heavy}
\sim {\rm M,AM,BM}$).
The calculated cross section
$\sigma (e^-e^- \rightarrow W^-W^-)$ for $\sqrt{s}=0.5(1)$ TeV and several
values A and B
as function of mass M and shown in Fig.1.
We can see from Eq.(15) that the cross section is larger for smaller
$m_{\rm heavy}$. The smallest value of $m_{\rm heavy}$ allowed in practice
is $\sim 100$ GeV the `see-saw' type models give the cross sections
$\sigma(e^-e^- \rightarrow W^-W^-)$ which are not experimentally interesting.

However the `see-saw' mechanism is not the only scenario which explains the
small
masses of the known neutrinos. There are models [11,12] where the relations
(14) do
not work and the mixing matrix elements can be large even for large masses of
the heavy neutrinos. In this class of models the smallness of masses of
the known neutrinos is
guaranteed by some special symmetry argument. There are then no simple
relations connecting $m_a$ with $K_{ae}$ and the mixing matrix elements can
be treated as independent parameters, bounded only by experimental data.
{}From existing experimental data only the sum
\begin{equation}
\sum\limits_{a=1}^{n_R} \mid K_{ae} \mid^2 \leq \kappa^2
\end{equation}
can be bounded:
$$\kappa^2=0.015 \;\;\;\; \eqno{ \rm (see\; Ref.[8]) }$$
or with the new LEP results ($m_t=170$ GeV and $m_H=200$ GeV)
$$\kappa^2=0.0054 \;\;\;\; \eqno{ \rm (see\; Ref.[9]).}$$
Let us calculate the total cross section
$\sigma(e^-e^- \rightarrow W^-W^-)$ for different number of right-handed
neutrinos
$n_R$. As was said the case $n_R=1$ is not interesting - the cross section
is very small. \\

$\bullet$ The $n_R=2$ case.

Let us denote the mass of the lightest heavy neutrino by $m_1=M$ and the mass
ratio
$m_2/m_1$ by A. Then, if we assume that $\eta_{CP}(N_1)=+i,\;
\eta_{CP}(N_2)=-i$ and denote $\delta=\Delta_{\rm light}/M <<1$, from Eqs.(13)
and
(17) (assuming the upper bound)  we have
\begin{eqnarray}
K_{N_1e}^2&=&\frac{A\kappa^2-\delta}{1+A}, \nonumber \\
K_{N_2e}^2&=&-\frac{\kappa^2-\delta}{1+A}.
\end{eqnarray}
The total cross section for mixing matrix elements (18) as function
of M for different ratios A is given in Fig.2. The largest value of the total
cross section is obtained for $\sqrt{s} \geq M$ (as discussed before)
and for $A \rightarrow \infty$. For very heavy second neutrino
$K_{N_2e} \rightarrow 0$ and the destructive interference in $R_{t(u)}$
functions
(Eq.(7)) between two neutrinos vanishes ($K_{N_1e} \rightarrow \kappa,\;
K_{N_2e} \rightarrow 0$). For $\kappa^2=0.0054$ the maximum of $\sigma_{\rm
tot}$
is obtained for $\sqrt{s}=0.5(1)$ TeV
$$\sigma_{\rm tot}({\rm max}) \simeq 2.3(10) fb\;\;\;{\rm for\;\;M}=400(700)
{\rm \;GeV}.$$
The cross section $\sigma_{\rm tot}$ depends crucially on the value of
$\kappa^2$. If we take for example the older value $\kappa^2=0.015$ we obtain
$$\sigma_{\rm tot}({\rm max}) \simeq 20(90) fb\;\;\;{\rm for\;\;M}=400(700)
{\rm \;GeV}.$$
The similar values of the $\sigma_{\rm tot}(max)$ were obtained in Ref.[5].
The assumption that the $\eta_{CP}$ of the heavy neutrinos is opposite
($\eta_{CP}(N_2)=-\eta_{CP}(N_1)=+i$) is equivalent to changing the sign of
$\delta$
and has no influence on the numerical values of the total cross section.
In the case of CP violation $K_{N_ie}$ are complex but still the bound
(17) is satisfied and $\sigma_{\rm tot}(max)$ is smaller than in the considered
case of the CP conservation. \\

$\bullet$ The $n_R=3$ case.

If we take
$$\eta_{CP}(N_1)=\eta_{CP}(N_2)=-\eta_{CP}(N_3)=+i$$
and  parametrize the heavy neutrino masses similarly as in the $n_R=2$ case
$$A=\frac{m_2}{M},\;\;B=\frac{m_3}{M},\;\;K_{N_1e}=x$$
we have from Eqs.(13) and (17)
\begin{eqnarray}
K_{N_2e}^2&=&\frac{B(\kappa^2-x^2)-x^2-\delta}{A+B}, \nonumber \\
K_{N_3e}^2&=&-\frac{A(\kappa^2-x^2)+x^2+\delta}{A+B},
\end{eqnarray}
where
$$0 \leq x^2 \leq \frac{B}{1+B}\kappa^2-\frac{\delta}{1+B}.$$
Now the cross section depends on four parameters M,A,B and x. As before, the
largest value of $\sigma_{\rm total}$ is obtained in the case when the
destructive
interference coming from the neutrino with $\eta_{CP}=-i$ disappears. It
happens for
$B \rightarrow \infty$ ($K_{N_3e} \rightarrow 0$).
The total cross sections
as functions of M for different values A and x ($0 \leq x \leq
\frac{B\kappa^2}{1+B}$) are
given in Fig.(3). We can see that the detailed behaviour of
$\sigma_{\rm tot}$
is now different than in the case $n_R=2$. For small values
of $x^2$, in particular,  the $\sigma_{\rm tot}({\rm max})$ is the same as in
the case $n_R=2$
and depends only on the value of $\kappa^2$. The different $\eta_{CP}$
configurations for neutrinos are obtained by the interchange $A \leftrightarrow
B$ and the sign change of $\delta$ and have no influence on $\sigma_{\rm tot}
({\rm max})$.

Finally, we check the influence of the unitary relation (9) on the
$\sigma_{\rm total}$ for the smaller values of $\sqrt{s}$. The unitary
constraints
begin to be important when $\sqrt{s} >> m_{\rm heavy}$ and cause the
$\sigma_{\rm tot} \rightarrow 0$ for $\sqrt{s} \rightarrow \infty$ (Eq.(8)).
The relation (9) is satisfied for heavy neutrinos if some of them have
opposite CP parities. To find how important the unitarity relation for
$\sqrt{s} \simeq m_{\rm heavy}$ is we assume that the CP parities of all
neutrinos are the same. In Fig.4 we present the behaviour of $\sigma_{\rm tot}$
as function of $\sqrt{s}$ if the unitarity relation is satisfied (dashed lines)
or
not satisfied (solid lines). The lines (h) show the results for three
right-handed
neutrinos with M=1 TeV A=B=2 and $x^2=\frac{\kappa^2}{2}$ (top of the (h) line
from Fig.3). We can see that for
$\sqrt{s}=1$ TeV the cross section where unitarity is not satisfied is
approximately one order of magnitude larger than the cross section which
satisfis the unitarity requirement.

The lines (f) show the results for M=700 GeV A=1 and B=100 ( $\sim$ top of the
(f) line
from Fig.3). The mass
of the third neutrino which causes the right or wrong unitarity behaviour
is large. The contribution of this very massive neutrino to
$\sigma_{\rm tot}$ is small and the difference between the right and wrong
unitarity behaviour is
visible only for very high energy ($\sqrt{s} \sim 10^4$ GeV). At the end we
would like to
stress that the dashed line (f) represents one of the most optimistic results
from
Fig.3.

\section{Conclusions}

We have calculated the total cross section for
the $e^-e^- \rightarrow W^-W^-$ process in the frame of the Standard Model
with additional right-handed neutrino singlets. The cross section resulting
from
the known light neutrino exchange is very small: $\sigma_{\rm tot} < 10^{-2}$
fb
if the laboratory limits for neutrino masses are taken or
$\sigma_{\rm tot} < 10^{-13}$ fb if astrophysical and cosmological bounds
are appropriate. The only chance to get a larger cross section is through
heavy Majorana neutrinos exchange with mass $M \geq M_Z$. If, however,
the small
masses of the existing neutrinos are explained by the `see-saw' mechanism,
the mixing matrix elements between electron and additional neutrinos
are small for large neutrino mass and the cross section is also small,
$\sigma_{\rm tot} < 10^{-3}$ fb (for $M \geq 100$ GeV).

In models where the `see-saw' mechanism is not employed to explain
the small masses of known neutrinos, the mixing matrix elements are
usually not connected with heavy neutrino mass. In such models the mixing
matrix
elements are free parameters which can be bounded from existing experimental
data.
Taking into account the new data from LEP the bounds on the mixing matrix
elements are such that the maximal cross section can be as large as
$\sigma_{\rm tot}({\rm max}) \simeq 2.3(10)$ fb for $\sqrt{s}=0.5(1)$ TeV.
Such large cross sections are possible only if the number of right-handed
neutrinos $n_R$ is greater than one ($n_R>1$) and the masses of heavy neutrinos
are different. The largest value
$\sigma_{\rm tot}({\rm max})$ is obtained for the energy $\sqrt{s}$ not very
different from the mass of the lightest heavy neutrino and only if the
neutrinos
with opposite CP parities are much heavier than the lighter ones. As the
present
experiments give bounds to the sum of the squares of moduli of the
mixing matrix elements
$\sum\limits_{a=1}^{n_R} \mid K_{ae} \mid^2,$ the value
$\sigma_{\rm tot}$
is independent of the number of the right-handed neutrinos. If CP is not
conserved
the $\sigma_{\rm tot}({\rm max})$ is smaller than in the case of CP
conservation.
The unitarity constraints can have a big influence on the value of the
$\sigma_{\rm tot}({\rm max})$ especially in the case where the destructive
interference between neutrinos with the opposite CP parities is large.

\section{Acknowledgement}
This work was supported by Polish Committee for Scientific Researches
under Grant No. PB 136/IF/95.

%%%%%%%%%%%%%%%%%%%%%%%%%%%%%%%%%%%%%%%%%%%%%%%%%%%%%%%%%%%%%%%%%%%%%%%%%%%%
\section*{References}
\newcounter{bban}
\begin{list}
{$[{\ \arabic {bban}\ }]$}{\usecounter{bban}\setlength{\rightmargin}{
\leftmargin}}
\item See e.g. Proc.of the Workshop on physics and experiments with linear
colliders (Saariselk$\ddot{a}$, Finland, September 1991),edited by
R.Orava,P.Eerola and
M.Nordberg (World Scientific, 1992) and Proc.of the Workshop on physics
and experiments with linear colliders (Waikoloa,Hawaii,April 1993),edited by
F.A.Harris,S.L.Olsen,S.Pakvasa and X.Tata (World Scientific, 1993).
\item J.F.Gunion and A.Tofighi-Niaki,Phys.Rev.D36,2671(1987) and
D38,1433(1988),
F.Cuypers,K.Ko{\l}odziej,O.Korakianitis and
R.R$\ddot{u}$kl,Phys.Lett.B325(1994)243.
\item `Inverse Neutrinoless Double $\beta$-Decay at the NLC?',T.Rizzo,
preprint ANL-HEP-CP-93-24.
\item J.Maalampi,A.Pietil$\ddot{a}$ and J.Vuori, Phys.Lett.B297(1992)327;
J.Gluza and M.Zra{\l}ek,hep-ph/9502284.
\item C.A.Heusch and P.Minkowski,Nucl.Phys.B416(1994)3 and `A strategy
for discovering heavy neutrinos',preprint BUTP-95/11,SCIPP 95/07.
%\item D.London,G.Belanger and J.N.Ng,Phys.Lett.B188,155(1987).
\item Particle Data Group,K.Hikasa et.al.,Phys.Rev.D,S1(1994).
\item O.Adriani et.al.Phys.Lett.B(1992)371.
\item E.Nardi,E.Roulet and D.Tommasini,Nucl.Phys.B386(1992);A.Ilakovic and
A.Pilaftsis Nucl.Phys.B437(1995)491.
\item A.Djoudi,J.Ng and T.G.Rizzo,`New Particles and Interactions',
SLAC-PUB-95-6772.
\item T.Yanagida,Prog.Theor.Phys.B135(1978)66; M.Gell-Mann,
P.Ramond and R.Slansky,in `Supergravity',edsP.Van Nieuwenhuizen
and D.Freedman (North-Holland,Amsterdam,1979)p.315.
\item D.Wyler and L.Wolfenstein,Nucl.Phys.B218(1983)205;R.N.Mohapatra and
J.W.F.Valle,Phys.Rev.D34(1986)1642;\newline
E.Witten,Nucl.Phys.B268(1986)79;
J.Bernabeu et al.,Phys.Lett.B187(1987)303;
J.L.Hewett and T.G.Rizzo,Phys.Rep.183
(1989)193;\newline
P.Langacker and D.London,Phys.Rev.D38(1988)907;E.Nardi,Phys.Rev.D48
(1993)3277;D.Tommasini et al.,`Non decoupling of heavy neutrinos and
leptons flavour violation' (hep-ph/9503228).
\item L.N.Chang,D.Ng and J.N.Ng,Phys.Rev.D50(1994)4589.
\item T.P.Cheng and L.F.Li,Phys.Rev.D44(1991)1502.
%\item E.Nardi,E.Roulet and D.Tommasini,Phys.Lett.B327(1994)319,B344(1995)225.
\end{list}
%%%%%%%%%%%%%%%%%%%%%%%%%%%%%%%%%%%%%%%%%%%%%%%%%%%%%%%%%%%%%%%%%%%%%%%%%%%%%%
\section*{Figure Captions}
\newcounter{bean}
\begin{list}
{\bf Fig.\arabic
{bean}}{\usecounter{bean}\setlength{\rightmargin}{\leftmargin}}
\item The cross section as function of the heavy neutrino mass for `classical'
see-saw models, where the mixing angles between light and heavy neutrinos are
proportional to the inverse of mass of the heavy neutrino. A=10, B=20
(Eq.(16)).
Solid (dashed) line is for the TLC (NLC) collider's energy.
\item  The cross section as function of the heavy neutrino masses for the
Standard Model
with two right-handed neutrinos.
Dashed (solid) lines are for $\sqrt{s}=0.5(1)$ TeV colliders.
The curves (a),(b),(c),(d) are for A=2,5,10,$10^5$ respectively (Eq.(18)).
\item  The cross section as function of the heavy neutrino masses for the
Standard Model
with three right-handed neutrinos for the TLC collider's energy.
The denotations are as follows (Eq.(19)):
(a):A=4,B=$\infty \;(10^5),x^2=0$;
(b):A=2,B=$\infty ,\;x^2=0$;
(c):A=2,B=$\infty ,\;x^2=0.002$;
(d):A=2,B=$\infty ,\;x^2=0.003$;
(e):A=2,B=$\infty ,\;x^2=0.004$;
(f):A=2,B=$\infty ,\;x^2=\kappa^2$;
(g):A=2,B=2,$\;x^2_{max}=\frac{2}{3}\kappa^2$;
(h):A=2,B=2,$\;x^2=\frac{\kappa^2}{2}$.
\item The influence of the unitarity constraints at the small energy limit for
the
RHS model with three right handed neutrinos (Eq.(19)). Lines denoted
by (f) are for A=1,B=100,M=700 GeV. Lines denoted by (h) are for
A=2,B=2,M=1000 GeV. Solid (dashed) lines are for real (complex)
coupling $K_{N_3e}$ (Eq.(19)).

\end{list}

\begin{table}
\begin{center}
\caption{
The reduced helicity amplitudes for
the $e^-e^- \rightarrow W^-W^-$ process in t- and u-channels.}
%$\left( \gamma= \frac{2M_W}{ \sqrt{s} },\; \beta= \sqrt{1- \gamma^2}
%\right) $ .}
\begin{tabular}{||c|c||c||c||} \hline \hline
$\lambda_1$
& $\lambda_2$
& $M_t$
& $M_u$  \\ \hline
 1 & 1 &
$ \gamma^2(1-c)$
   & $\gamma^2(1+c)$   \\
&&& \\
-1 & -1 & $ \gamma^2(1+c)$
  &  $ \gamma^2(1-c)$
  \\ \cline{1-4}
 1 & 0  &
$ -\gamma(1-\beta)$
 &
$ \gamma(1-\beta)$ \\
&&& \\
 -1 & 0 & $ \gamma(1+\beta)$
& $ -\gamma(1+\beta)$
\\ \cline{1-4}
 0 & 1  &
$ -\gamma(1-\beta)$ &
$ \gamma(1-\beta)$  \\
&&& \\
 0 & -1 & $ \gamma(1+\beta)$
& $ -\gamma(1+\beta)$
\\ \cline{1-4}
 1 & -1 & 0 & 0 \\
 -1 & 1 & 0 & 0 \\  \cline{1-4}
&&& \\
 0 & 0 & $-\left( 1+\beta^2 - 2c \beta \right) $
 &
$- \left( 1+\beta^2 + 2c \beta \right) $
\\ \hline \hline
\end{tabular}
\end{center}
\end{table}
\newpage
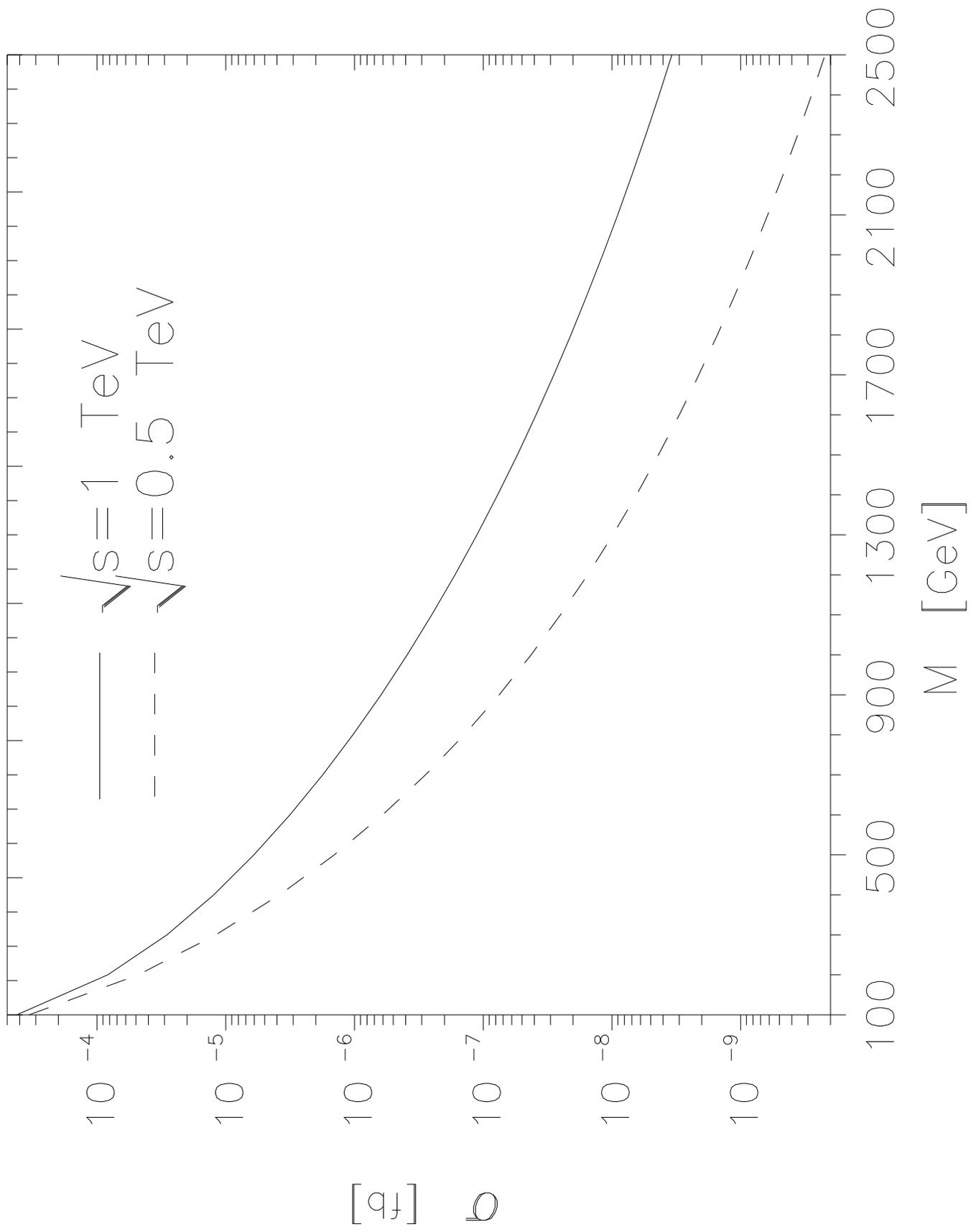
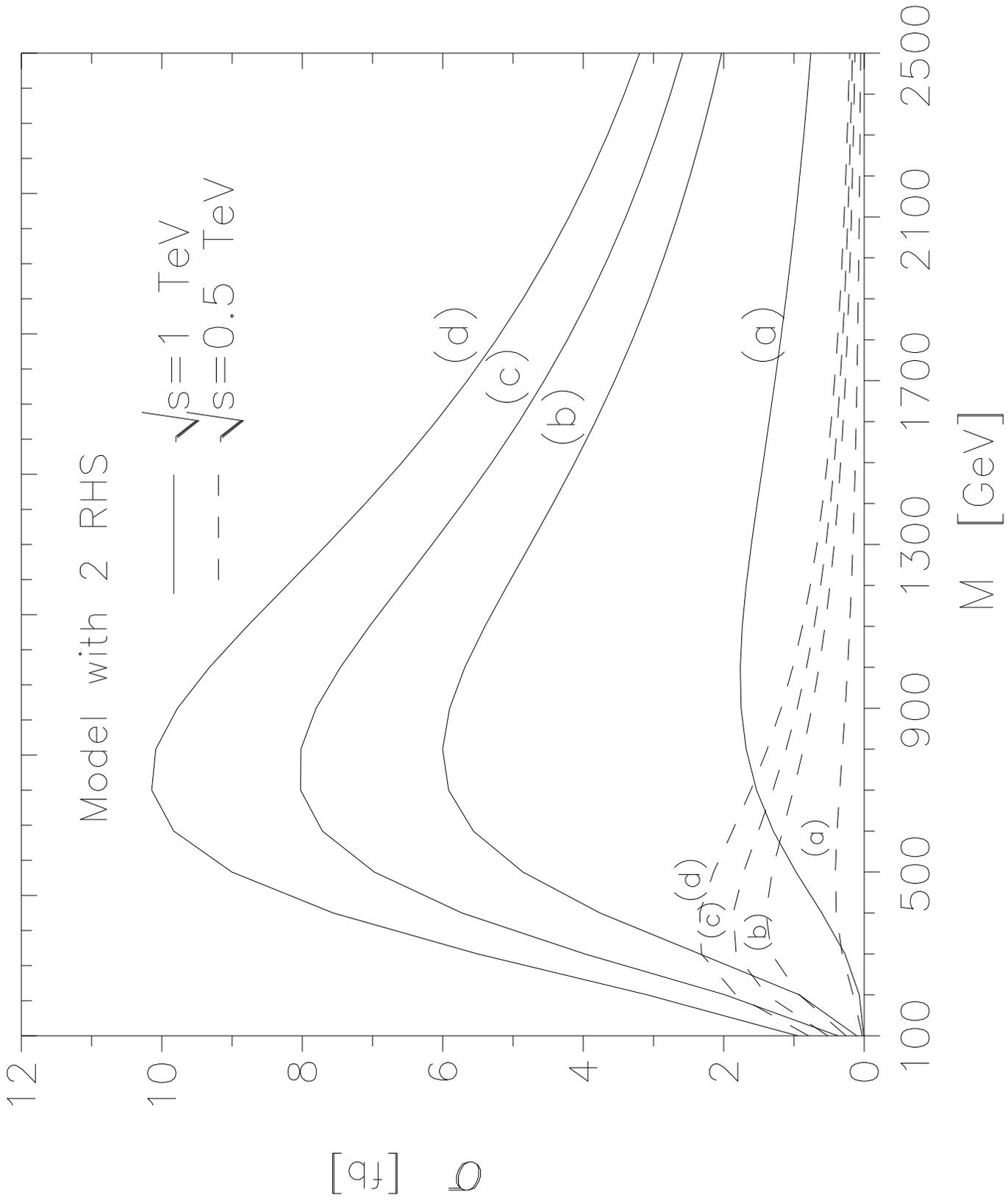
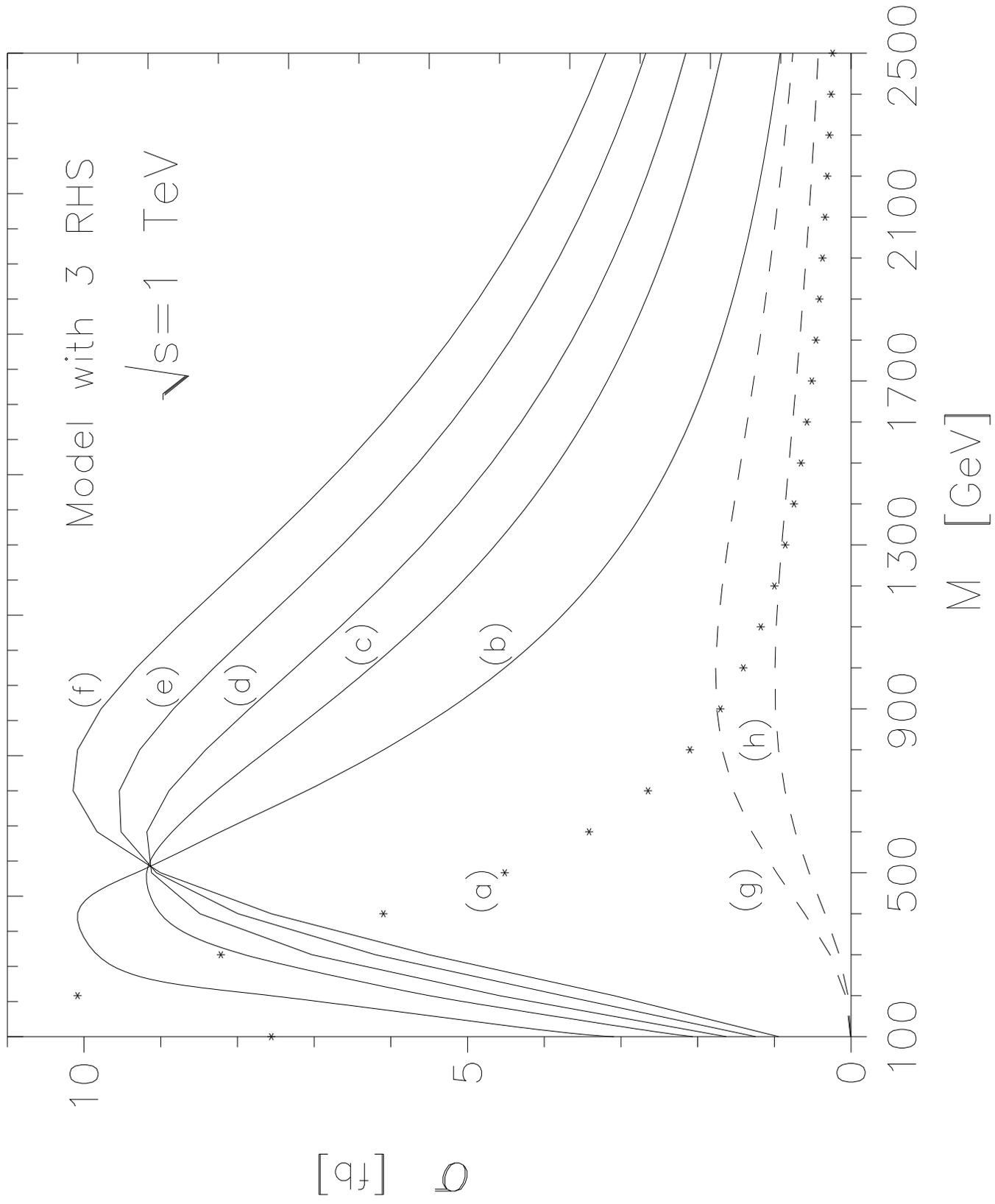
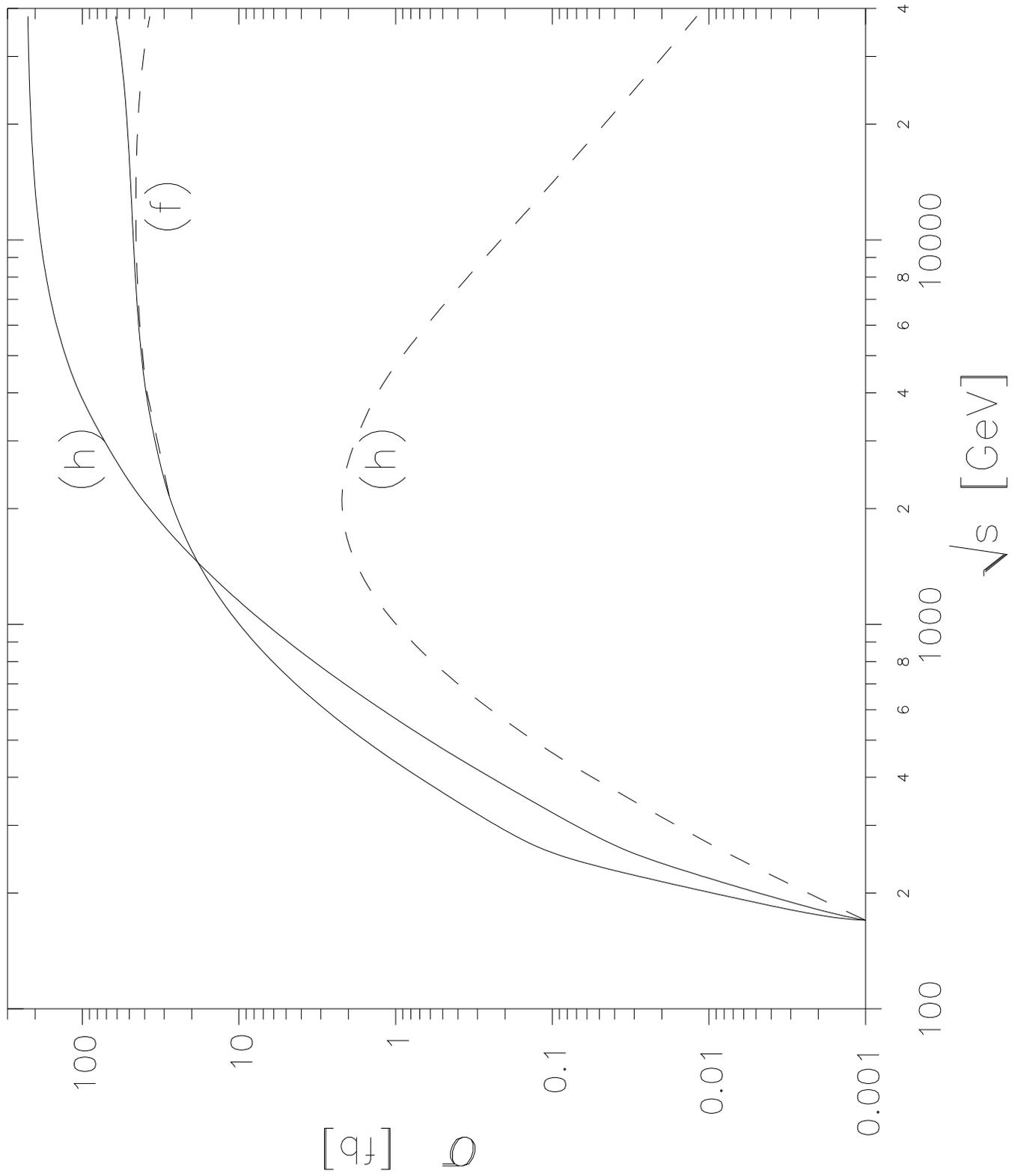
\end{document}